\documentclass[a4paper,11pt]{article}
\usepackage{pos}
\usepackage{comment}

\title{Event-by-event net-charge fluctuations in pp, OO, Ne--Ne, and Pb--Pb collisions with ALICE at the LHC}
\ShortTitle{Net-charge fluctuations in pp, OO, Ne--Ne, and Pb--Pb collisions}

\author*[a]{Nida Malik}

\onbehalf{for the ALICE Collaboration}

\affiliation[a]{Department of Physics, Aligarh Muslim University\\
  Aligarh-202002, India}

\emailAdd{nida.malik@cern.ch, nidamalik7723@gmail.com}

\abstract{

\noindent Investigations involving the event-by-event fluctuations of conserved quantities, like net charge, net baryon number, and strangeness in heavy-ion collisions, provide insights into the properties of quark–gluon plasma (QGP) and the phase diagram of strongly interacting matter. Event-by-event fluctuations of net electric charge in {\it pp} collisions at $\sqrt{s} = 13$ and $13.6$ TeV and for {\it OO, Ne--Ne, and Pb--Pb} collisions at $\sqrt{s_{\text{NN}}} = 5.36$ TeV are studied by analysing the data collected by the ALICE detector at CERN Large Hadron Collider.
\noindent The analysis is carried out in terms of the variable  $\nu_{\text{dyn}}[+,-]$, chosen for its robustness against the detector efficiency losses. 
The observed dependence of $\nu_{\rm dyn}[+,-]$ on the charged-particle density shows a smooth, energy-independent decrease in the magnitude of $\nu_{\rm dyn}[+,-]$ from smaller to larger collision systems.

\noindent The negative values of $\nu_{\text{dyn}}[+,-]$ indicate a dominance of correlations between oppositely charged particle pairs compared with those arising from like-sign charge pairs. Since $\nu_{\text{dyn}}[+,-]$ is known to have an intrinsic dependence on multiplicity, an appropriate scaling is applied. The variable is therefore scaled with the mean multiplicities,$\frac{1}{\frac{1}{\langle N_+ \rangle} + \frac{1}{\langle N_- \rangle}}$. The dependence of the scaled values of $\nu_{\text{dyn}}[+,-]$  on charged particle density is also looked into.  The results are compared with model calculations from PYTHIA, EPOS4, AMPT and HIJING.

}

\FullConference{14th Edition of the Large Hadron Collider Physics (LHCP2026)\\
18-22 May 2026\\
Paris, France\\}

\begin{document} 
\maketitle

\section{Introduction}

\noindent It is commonly believed that the event-by-event (ebe) fluctuations observed in hadronic and nucleus-nucleus (AA) collisions are primarily connected to the idea that the correlation and fluctuations of dynamical origin are associated to the critical phenomena of the phase transition and lead to local and global differences between the events produced under similar initial conditions \cite{DeWolf1996,Ahmad2018}. Ebe fluctuations in conserved quantities, like strangeness, electric charge and baryon number would help estimate the degree of equilibration and criticality of the measured system~\cite{Sharma2015}. It has been argued that the phase transition from QGP to normal hadronic matter is an entropy-conserving process~\cite{Gosh2017} and hence the fluctuations in net electric charge would be reduced significantly in the final state in comparison to what is expected from the hadron gas (HG) system~\cite{Jeon1999,Heiselberg2001}. This is expected because the magnitude of charge fluctuations is proportional to the square of the number of charges present in the system which depends on the state from which charges originate. In a system passing through the QGP phase, quarks are the charge carriers, whereas in the case of HG, the charge carriers are hadrons. This suggests that the charge fluctuations observed in the case of QGP with fractional charges would be smaller than those in HG with integral charges~\cite{Sharma2015,Jeon2000,Bleicher2000}.

\noindent Experimentally, net-charge fluctuations are affected by event-by-event variations in the collision geometry, which introduce volume fluctuations. To reduce such effects, the observable $\nu_{\mathrm{dyn}}[+,-]$ is commonly employed~\cite{Pruneau2002,STAR2003,ALICE2013}. It is defined as
\begin{equation}
\nu_{\mathrm{dyn}}[+,-] = \frac{\langle N_{+}(N_{+}-1) \rangle}{\langle N_{+} \rangle^{2}} +
\frac{\langle N_{-}(N_{-}-1) \rangle}{\langle N_{-} \rangle^{2}} - 2\frac{\langle N_{+}N_{-} \rangle}
{\langle N_{+} \rangle \langle N_{-} \rangle},
\label{eq:nudyn}
\end{equation}
where $N_{+}$ and $N_{-}$ denote the multiplicities of positively and negatively charged particles, respectively, while the angular brackets denote event-averaged quantities. The observable $\nu_{\mathrm{dyn}}[+,-]$ isolates dynamical correlations between charged particles, quantifies deviations from independent particle emission ($\nu_{\mathrm{dyn}}[+,-]=0$), and is largely insensitive to detector efficiency. Since electric charge is conserved, one expects $\nu_{\mathrm{dyn}}[+,-]$ to be negative, i.e., the production of positively and negatively charged particles is correlated.

\noindent Although $\nu_{\mathrm{dyn}}[+,-]$ is independent of fluctuations in the number of particle sources, it exhibits an inverse dependence on the system size and is, therefore, neither an intensive nor an extensive quantity. To address this limitation, observables that are insensitive to both the system volume and its fluctuations on the ebe basis have been introduced~\cite{ref5,Khan2022}. Such observables, commonly referred to as strongly intensive quantities, provide a unified framework for fluctuation studies across different collision systems.

\noindent One of the strongly intensive measures of net-charge fluctuations is defined as
\begin{equation}
\Sigma[N_+,N_-] = \frac{1}{\langle N_+ \rangle+\langle N_- \rangle}\left[\langle N_+ \rangle\omega_- +
\langle N_- \rangle\omega_+ - 2\mathrm{Cov}(N_+,N_-) \right],
\label{eq:sigma}
\end{equation}
where $\omega_+$ ($\omega_-$) is the scaled variance of positively (negatively) charged particles, defined as
\begin{equation}
\omega_\pm = \frac{\langle N_\pm^2\rangle-\langle N_\pm\rangle^2} {\langle N_\pm\rangle}.
\label{eq:omega}
\end{equation}
The quantity $\Sigma$ is dimensionless and normalized such that $\Sigma[N_+,N_-]=\Sigma[N_-,N_+]=1$ for statistically independent particle emission~\cite{ref5}.

\noindent At {\footnotesize LHC} energies, where the condition $\langle N_+\rangle \approx \langle N_-\rangle$ is well satisfied, the strongly intensive quantity $\Sigma$ can be related to $\nu_{\mathrm{dyn}}[+,-]$ through the following equation:
\begin{equation}
\Sigma[N_+,N_-]-1 = \frac{\nu_{\mathrm{dyn}}[+,-]} {\left(\frac{1}{\langle N_+\rangle}+ \frac{1}{\langle N_-\rangle} \right)}.
\label{eq:sigma_nudyn}
\end{equation}
\noindent In the present analysis, the scaled form of $\nu_{\mathrm{dyn}}[+,-]$, as defined by Eq.~\ref{eq:sigma_nudyn}, is employed to account for its intrinsic multiplicity dependence. Unlike $\nu_{\mathrm{dyn}}[+,-]$, the scaled quantity requires efficiency corrections to the mean multiplicities, $\langle N_+\rangle$ and $\langle N_-\rangle$.

\section{Analysis Details}

\noindent 
In the present study, data involving various collision systems (small, intermediate as well as large) at various energies collected by the ALICE detector at CERN at the LHC are used. These are listed in Table~\ref{tab:datasets}. These datasets would allow studying beam energy and system size dependence of net-charge fluctuations.

\begin{table}[h!]
\centering
\caption{Summary of datasets used in the analysis.}
\label{tab:datasets}
\begin{tabular}{|l|c|c|}
\hline
\textbf{Collision System} & \textbf{$\sqrt{s_{\mathrm{NN}}}$ (TeV)} & \textbf{Number of events (Million)} \\
\hline
pp (Run2 converted)        & 13   &  740  \\
pp (Run3)                  & 13.6 & 2200  \\
OO (Run3)                 & 5.36 & 581 \\
Ne--Ne (Run3)              & 5.36 & 498 \\
Pb--Pb (Run3)              & 5.36 & 46.4  \\
\hline
\end{tabular}
\end{table}
\noindent For {\it pp} collisions, multiplicity is estimated using the FT0M estimator in Run~3 and the V0M estimator in Run~2, while the FT0C estimator is used to determine the centrality of {\it OO}, {\it Ne--Ne}, and {\it Pb--Pb} collisions.
 The primary vertex, $V_z$ is reconstructed using the ITS and TPC, with $|V_z|<10$~cm. Minimum-bias events are selected using the \texttt{sel8} trigger for Run~3 and the \texttt{sel7} trigger for Run~2, with additional pile-up rejection criteria.

\noindent Charged-particle tracks are reconstructed using the ITS as well as TPC and selected within $0.2 < p_{\mathrm{T}} < 5.0$~GeV/$c$ and $|\eta|<0.8$. The mean positive- and negative-charged particle multiplicities are corrected for efficiency using Monte Carlo simulations. Statistical uncertainties are evaluated using the subsample method and systematic uncertainties are estimated by varying the event and track selection criteria, including $|V_z|$, TPC crossed rows, $\chi^2$ per ITS and TPC cluster, DCA$_{xy}$, DCA$_z$, and Monte Carlo non-closure is also included.

\section{Results}

\begin{figure}
\centering
\includegraphics[width=0.65\linewidth]{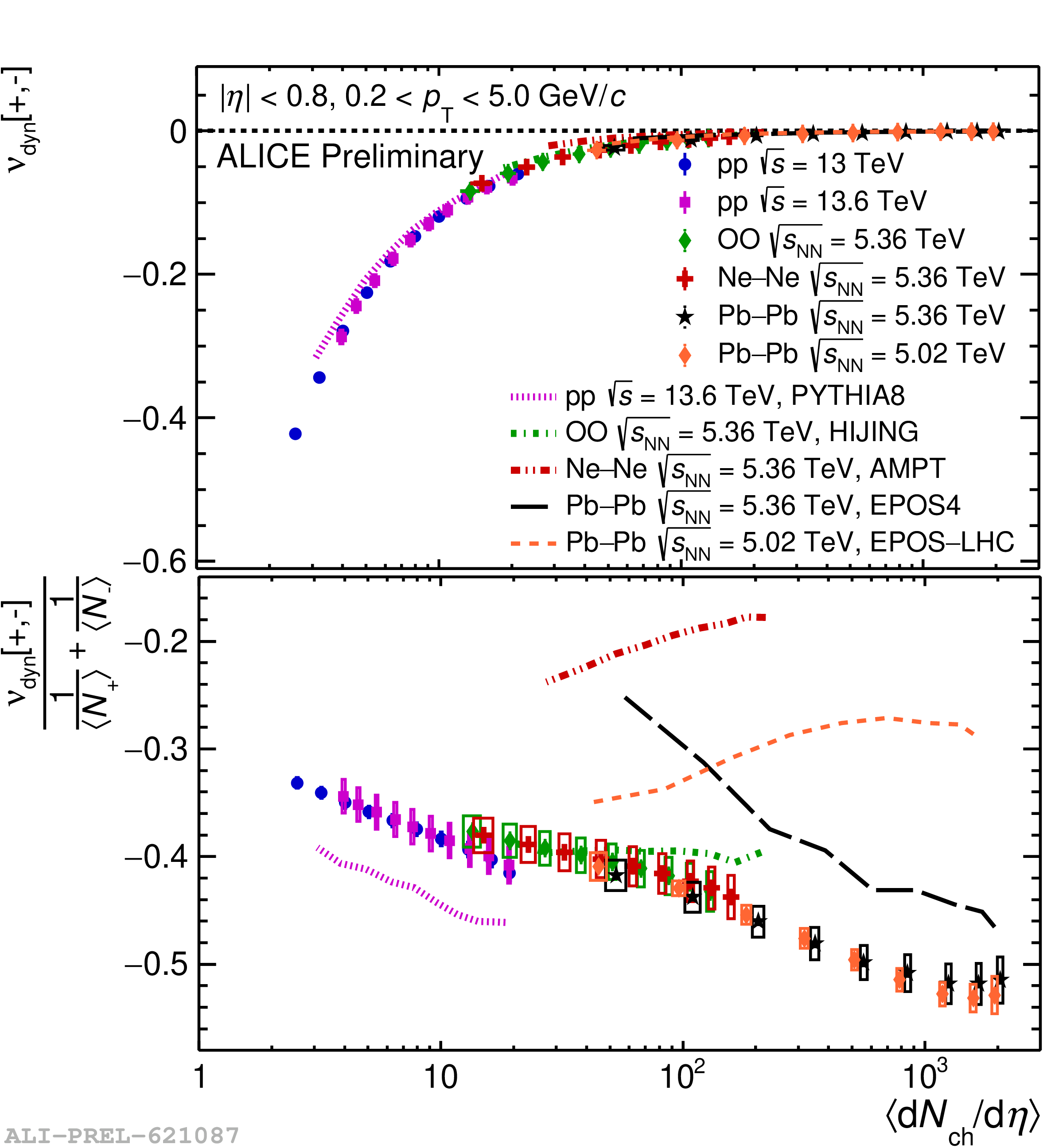}
\caption{$\nu_{\mathrm{dyn}}[+,-]$ (top panel) and its scaled form (bottom panel) as a function of the charged-particle density $\langle dN_{\mathrm{ch}}/d\eta\rangle$ for {\it pp}, {\it OO}, {\it Ne--Ne}, and {\it Pb--Pb} collisions.}
\label{fig:scaled_nudyn}
\end{figure}

\noindent To investigate the evolution of net-charge fluctuations across different collision systems, \(\nu_{\mathrm{dyn}}[+,-]\) is studied as a function of the charged-particle density \(\langle dN_{\mathrm{ch}}/d\eta\rangle\), providing a common multiplicity scale. For each multiplicity/centrality class, \(\nu_{\mathrm{dyn}}[+,-]\) is calculated independently, and the corresponding \(\langle dN_{\mathrm{ch}}/d\eta\rangle\) values are obtained from available ALICE results for the respective multiplicity/centrality classes. For Pb--Pb collisions at \(\sqrt{s_{\mathrm{NN}}}=5.36\) TeV, the charged-particle density values are taken from Ref.~\cite{ALICE2025dNch}. The dependence of \(\nu_{\mathrm{dyn}}[+,-]\) on \(\langle dN_{\mathrm{ch}}/d\eta\rangle\) is shown in the top panel of Fig.~\ref{fig:scaled_nudyn}.  For all collision systems, $\nu_{\mathrm{dyn}}[+,-]$ is observed to remain predominantly negative over the entire multiplicity range, reflecting correlated production of oppositely charged particles, with contributions from mechanisms such as local charge conservation and resonance decays. The magnitude of net-charge fluctuations decreases with increasing $\langle dN_{\mathrm{ch}}/d\eta\rangle$ and tends to zero at high multiplicities, indicating a dilution of these correlations with increasing charged particle density. A smooth evolution is observed from small ({\it pp}) to intermediate ({\it OO}, {\it Ne--Ne}) and large ({\it Pb--Pb}) collision systems, suggesting that the observed behavior is primarily governed by the charged-particle multiplicity rather than the collision system size.\\

\noindent The results are compared with model calculations from PYTHIA~8 for {\it pp} collisions, HIJING for {\it OO}, AMPT for {\it Ne--Ne}, and EPOS4 for {\it Pb--Pb} collisions at $\sqrt{s_{\mathrm{NN}}}=5.36$~TeV. The model calculations exhibit different levels of agreement with the measured $\nu_{\mathrm{dyn}}[+,-]$ dependence across the collision systems. The preliminarily approved results for Pb--Pb collisions at \(\sqrt{s_{\mathrm{NN}}}=5.02\) TeV from ALICE Run 2, reported in Ref.~\cite{Khan2022}, along with EPOS-LHC predictions at the same collision energy, have also been included for comparison.\\

\noindent To account for the intrinsic multiplicity dependence of $\nu_{\mathrm{dyn}}[+,-]$, a scaled observable is constructed as
\begin{equation}
\frac{\nu_{\mathrm{dyn}}[+,-]} {\left(\frac{1}{\langle N_+\rangle}+ \frac{1}{\langle N_-\rangle} \right)}
\label{eq:scaled_nudyn}
\end{equation}
The mean multiplicities used in the above relation are efficiency-corrected.\\

\noindent Variations of scaled values of $\nu_{\mathrm{dyn}}[+,-]$ with $\langle dN_{\mathrm{ch}}/d\eta\rangle$ are shown in the bottom panel of Fig.~\ref{fig:scaled_nudyn}. Unlike $\nu_{\mathrm{dyn}}[+,-]$, which approaches zero with increasing multiplicity, the scaled observable becomes more negative with increasing charged-particle density. This behavior demonstrates a systematic evolution of the net-charge fluctuations with multiplicity after removing the intrinsic multiplicity dependence of $\nu_{\mathrm{dyn}}[+,-]$.

\noindent The observed behavior may be influenced by global charge conservation and resonance decays coupled with radial flow. Comparison with model calculations provides further insight into the relative contributions of these effects and their ability to describe the multiplicity dependence across different collision systems.

\section{Conclusions}

\noindent Event-by-event net-charge fluctuations have been studied using the observable $\nu_{\mathrm{dyn}}[+,-]$ in {\it pp} collisions at $\sqrt{s}=13$ and $13.6$~TeV and in {\it OO}, {\it Ne--Ne}, and {\it Pb--Pb} collisions at $\sqrt{s_{\mathrm{NN}}}=5.36$~TeV. The observable remains negative over the measured charged-particle density range for all collision systems, indicating the dominance of correlations between oppositely charged particles. Its magnitude decreases with increasing $\langle dN_{\mathrm{ch}}/d\eta\rangle$ and approaches zero toward higher multiplicities, showing a smooth evolution from small to large collision systems.

\noindent To account for the intrinsic multiplicity dependence of $\nu_{\mathrm{dyn}}[+,-]$, a scaled form related to the strongly intensive observable $\Sigma$ is studied. The scaled observable becomes more negative with increasing charged-particle density across the collision systems, indicating a systematic evolution of opposite-charge correlations with multiplicity. The observed correlations can have contributions from mechanisms such as global charge conservation and resonance decays coupled with radial flow.

\noindent Comparisons with model calculations show different levels of agreement with the data. The measurements provide a systematic study of net-charge fluctuations and charge correlations over a broad range of collision systems and charged-particle multiplicities at LHC energies.

\end{document}